\documentclass{ws-procs9x6}

\def\citebk#1{[\hspace{0.9mm}\raisebox{-1.85mm}[0mm][0mm]
  {\Large\cite{#1}}\hspace{-0.1mm}]}

\def\citebkcap#1{[\hspace{0.8mm}\raisebox{-1.5mm}[0mm][0mm]
  {\large\cite{#1}}\hspace{-0.2mm}]}

\begin{document}

\title{\vspace{-3mm}
\rightline{\small IFUP-TH 2002/28}
\vspace{8mm}
Finite-temperature behavior of the (2+1)D Georgi-Glashow model
with and without quarks}

\author{Dmitri Antonov}

\address{INFN-Sezione di Pisa, Universit\'a degli studi di Pisa,
Dipartimento di Fisica, Via Buonarroti, 2 - Ed. B - 56127 Pisa, 
Italy\\
E-mail: antonov@df.unipi.it}


\maketitle

\abstracts{(2+1)-dimensional Georgi-Glashow model and its $SU(N)$-generalization are explored 
at nonzero temperatures and in the
regime when the Higgs boson is not infinitely heavy.
The finiteness of the Higgs-boson mass leads to various novel effects. Those include
the appearance of two separate phase transitions and of the upper bound on the 
parameter of the weak-coupling approximation,
necessary to maintain the stochasticity of the Higgs vacuum.
The modification of the finite-temperature behavior of the 
model emerging due to the introduction of massless
quarks is also discussed.}

\section{Introduction. The model}
Since the second half of the seventies~\citebk{1},
(2+1)-dimensional Georgi-Glashow model is known as
an example of the theory allowing for an analytic description
of confinement. However, confinement in the Georgi-Glashow model is typically discussed
in the limit of infinitely large Higgs-boson mass, when the
model is reduced to compact QED. In the present talk, we shall review various novel effects stemming 
from the finiteness of that mass. The main emphasis in this discussion will be 
payed to the modification of the finite-temperature properties of the Georgi-Glashow model, as well as 
of its $SU(N)$-generalization.

The Euclidean action of the (2+1)D Georgi-Glashow 
model reads

\begin{equation}
\label{GG}
S=\int d^3x\left[\frac{1}{4g^2}\left(F_{\mu\nu}^a\right)^2+
\frac12\left(D_\mu\Phi^a\right)^2+\frac{\lambda}{4}\left(
\left(\Phi^a\right)^2-\eta^2\right)^2\right].
\end{equation}
Here, the Higgs field $\Phi^a$ transforms by the adjoint representation, 
$D_\mu\Phi^a\equiv\partial_\mu\Phi^a+\varepsilon^{abc}A_\mu^b
\Phi^c$. Next, $\lambda$ is the Higgs coupling constant of dimensionality [mass], 
$\eta$ is the Higgs v.e.v. of dimensionality $[{\rm mass}]^{1/2}$, and 
$g$ is the electric coupling constant of the same dimensionality.
At the one-loop level, the sector of the theory~(\ref{GG}) containing 
dual photons and Higgs bosons is represented by the following partition function~\citebk{dietz}:

$$
{\mathcal Z}=1+\sum\limits_{N=1}^{\infty}\frac{\zeta^N}{N!}\left(\prod\limits_{i=1}^{N}\int d^3z_i
\sum\limits_{q_i=\pm 1}^{}\right)\times
$$

\begin{equation}
\label{pf}
\times\exp\left[-\frac{g_m^2}{8\pi}
\sum\limits_{{a,b=1\atop a\ne b}}^{N}\left(\frac{q_aq_b}{|{\bf z}_a-{\bf z}_b|}-
\frac{{\rm e}^{-m_H|{\bf z}_a-{\bf z}_b|}}{|{\bf z}_a-{\bf z}_b|}\right)\right]\equiv
\int {\mathcal D}\chi{\mathcal D}\psi {\rm e}^{-S},
\end{equation}
where

$$
S=\int d^3x\left[\frac12(\partial_\mu\chi)^2+\frac12(\partial_\mu\psi)^2
+\frac{m_H^2}{2}\psi^2-2\zeta{\rm e}^{g_m\psi}\cos(g_m\chi)\right]\equiv
$$

\begin{equation}
\label{1}
\equiv\int d^3x{\mathcal L}[\chi,\psi|g_m,\zeta].
\end{equation}
The partition function~(\ref{pf}) describes the grand canonical ensemble of monopoles with the 
account for their Higgs-mediated interaction.
In Eqs.~(\ref{pf}) and~(\ref{1}), $\chi$ is the dual-photon field, and the field $\psi$ accounts for the Higgs field,
whose mass reads $m_H=\eta\sqrt{2\lambda}$. Note that from Eq.~(\ref{pf}) it is straightforward to deduce that when $m_H$
formally tends to infinity, one arrives at the conventional sine-Gordon theory of the 
dual-photon field~\citebk{1} describing the compact-QED limit of the model.
Next, in the above equations, $g_m$ stands for the magnetic coupling constant 
related to the electric one as $g_mg=4\pi$, and 
the monopole fugacity $\zeta$ has the form
$\zeta=\frac{m_W^{7/2}}{g}\delta\left(\frac{\lambda}{g^2}\right)
{\rm e}^{-4\pi m_W\epsilon/g^2}$.
In this formula, $m_W=g\eta$ is the W-boson mass, 
and $\epsilon=\epsilon(\lambda/g^2)$ is a certain monotonic, slowly 
varying function, $\epsilon\ge 1$, $\epsilon(0)=1$~\citebk{bps},  
$\epsilon(\infty)\simeq 1.787$~\citebk{kirk}.
As far as the function $\delta$ is concerned, 
it is determined by the loop corrections. In what follows, we shall work 
in the standard weak-coupling regime $g^2\ll m_W$, which parallels the requirement 
that $\eta$ should be large enough to ensure the spontaneous symmetry breaking from 
$SU(2)$ to $U(1)$. The W-boson mass will thus play the role of the UV cutoff in the further analysis.

\section{The model at finite temperature beyond the compact-QED limit}

In the discussion of finite-temperature properties of the Georgi-Glashow model in the
present and next Sections, we shall mainly follow Ref.~\citebk{plb}.
At finite temperature $T\equiv1/\beta$, one should supply the fields $\chi$ and $\psi$
with the periodic boundary conditions in the temporal direction, with the period equal to
$\beta$. Because of that, the lines of magnetic field emitted by a monopole cannot cross
the boundary of the one-period region and consequently, at the distances larger than $\beta$,  
should go almost parallel to this boundary, approaching it. 
Therefore, monopoles separated by such distances
interact via the 2D Coulomb potential, rather than the 3D one. Since the average distance
between monopoles in the plasma is of the order $\zeta^{-1/3}$, we see that at $T\gtrsim\zeta^{1/3}$,
the monopole ensemble becomes two-dimensional. Owing to the fact that $\zeta$ is exponentially 
small in the weak-coupling regime under discussion, the idea of dimensional reduction is perfectly applicable
at the temperatures of the order of the critical temperature of the Berezinsky-Kosterlitz-Thouless (BKT)
phase transition~\citebk{bkt} (see e.g. Ref.~\citebk{rev} for a review) in the monopole plasma, 
which is equal to $g^2/2\pi$~\citebk{2}~\footnote{Note that due to the $T$-dependence of the strength of the 
monopole-antimonopole interaction, which is a consequence of the dimensional reduction, the BKT 
phase transition in the 3D Georgi-Glashow model is inverse with respect to the standard one of the 2D 
XY model. Namely, monopoles exist in the plasma phase at the temperatures below the BKT critical one
and in the molecular phase otherwise.}. Up to exponentially small corrections, this temperature is 
unaffected by the finiteness of the Higgs-boson mass. This can be seen from the expression for the 
mean squared separation in the monopole-antimonopole molecule, 

$$
\left<L^2\right>=\frac{\int\limits_{|{\bf x}|>m_W^{-1}}^{} d^2{\bf x}|{\bf x}|^{2-\frac{8\pi T}{g^2}}\exp
\left[\frac{4\pi T}{g^2}K_0\left(m_H|{\bf x}|
\right)\right]}{\int\limits_{|{\bf x}|>m_W^{-1}}^{} 
d^2{\bf x}|{\bf x}|^{-\frac{8\pi T}{g^2}}\exp\left[\frac{4\pi T}{g^2}K_0\left(m_H|{\bf x}|
\right)\right]},$$
where $K_0$ denotes the modified Bessel function.
Disregarding the exponential factors in the numerator and denominator of this equation, we obtain 
$\left<L^2\right>\simeq\frac{4\pi T-g^2}{2m_W^2\left(2\pi T-g^2\right)}$,
that yields the above-mentioned value of the BKT critical temperature $g^2/2\pi$. 
Besides that, it is straightforward to see that in the weak-coupling 
regime under study, the value of $\sqrt{\left<L^2\right>}$ is exponentially smaller than the characteristic distance in the 
monopole plasma, $\zeta^{-1/3}$, i.e., molecules are very small-sized with respect to that distance.

The factor $\beta$ at the 
action of the dimensionally-reduced theory, $S_{{\rm d.-r.}}=\beta\int d^2x{\mathcal L}[\chi,\psi|g_m,\zeta]$,
can be removed [and this action can be cast to the original form of eq.~(\ref{1}) with the substitution $d^3x\to d^2x$]
by the obvious rescaling: 
$S_{{\rm d.-r.}}=\int d^2x{\mathcal L}\left[\chi^{\rm new},\psi^{\rm new}|\sqrt{K},\beta\zeta\right]$.
Here, $K\equiv g_m^2T$, 
$\chi^{\rm new}=\sqrt{\beta}\chi$, $\psi^{\rm new}=\sqrt{\beta}\psi$, and in what follows
we shall denote for brevity $\chi^{\rm new}$ and $\psi^{\rm new}$ simply as $\chi$ and $\psi$, respectively.
Averaging then over the field $\psi$ with the use of the cumulant expansion we arrive at the 
following action:

$$
S_{{\rm d.-r.}}\simeq\int d^2x\left[\frac12(\nabla\chi)^2-2\xi\cos\left(g_m\sqrt{T}\chi\right)\right]-
$$

\begin{equation}
\label{2}
-2\xi^2\int d^2xd^2y\cos\left(\sqrt{K}\chi({\bf x})\right){\mathcal K}^{(2)}({\bf x}-{\bf y})
\cos\left(\sqrt{K}\chi({\bf y})\right).
\end{equation}
In this expression, we have disregarded all the cumulants higher 
than the quadratic one, and the limits of applicability of 
this so-called bilocal approximation will be discussed below. 
Further, in Eq.~(\ref{2}), ${\mathcal K}^{(2)}({\bf x})\equiv{\rm e}^{KD_{m_H}^{(2)}({\bf x})}-1$, where  
$D_{m_H}^{(2)}({\bf x})\equiv K_0(m_H|{\bf x}|)/2\pi$ is the 2D Yukawa propagator, and 
$\xi\equiv\beta\zeta{\rm e}^{\frac{K}{2}D_{m_H}^{(2)}(0)}$
denotes the monopole fugacity modified by the interaction of monopoles via the Higgs field.
Clearly, in the compact-QED limit (when $m_H$ formally tends to infinity) $D_{m_H}^{(2)}(0)$, being equal to 
$\int\frac{d^2p}{(2\pi)^2}\frac{1}{p^2+m_H^2}$, vanishes already before doing the integration, and $\xi\to\beta\zeta$,
as it should be. In the general case, when the mass of the Higgs field is moderate
and does not exceed $m_W$, we obtain

$$\xi\propto\exp\left[-\frac{4\pi}{g^2}\left(m_W\epsilon+T\ln\left(\frac{{\rm e}^{\gamma}}{2}c\right)
\right)\right].$$
Here, we have introduced the notation $c\equiv m_H/m_W$, $c<1$, and $\gamma\simeq 0.577$ is the 
Euler constant, so that $\frac{{\rm e}^{\gamma}}{2}\simeq 0.89<1$. We see that 
the modified fugacity remains exponentially small, provided that 

\begin{equation}
\label{3}
T<-\frac{m_W\epsilon}{\ln\left(\frac{{\rm e}^{\gamma}}{2}c\right)}.
\end{equation}

This constraint should be updated by another one, which would provide the convergence
of the cumulant expansion applied in course of the average over $\psi$. Were the cumulant expansion 
divergent, this fact would signal that the Higgs vacuum loses its normal stochastic properties and 
becomes a coherent one.
In order to get the 
new constraint, notice that the parameter of the cumulant expansion reads 
$\xi I^{(2)}$, where $I^{(2)}\equiv\int d^2x{\mathcal K}^{(2)}({\bf x})$.
Evaluation of the integral $I^{(2)}$ yields~\citebk{plb}:

\begin{equation}
\label{I}
I^{(2)}\simeq\frac{2\pi}{m_H^2}\left[\frac12\left(c^2-1+\left(\frac{2}{{\rm e}^{\gamma}}
\right)^{\frac{8\pi T}{g^2}}\frac{1-c^{2-
\frac{8\pi T}{g^2}}}{1-\frac{4\pi T}{g^2}}\right)
+{\rm e}^{\frac{a}{{\rm e}}}-1+\frac{a}{{\rm e}}\right].
\end{equation}
(Note that at $T\to g^2/4\pi$, $\frac{1-c^{2-\frac{8\pi T}{g^2}}}{1-\frac{4\pi T}{g^2}}\to
-2\ln c$, i.e., $I^{(2)}$ remains finite.)
In the derivation of this expression, 
the parameter $a\equiv4\pi\sqrt{2\pi}T/g^2$ was assumed to be of the order of unity.
That is because the temperatures we are working at are of the order of the BKT critical one,
$g^2/2\pi$. 
Due to the exponential term in Eq.~(\ref{I}),
the violation of the cumulant expansion may occur at high enough temperatures [that parallels the above-obtained 
constraint~(\ref{3})].
The most essential, exponential, part of the parameter
of the cumulant expansion thus reads

$$\xi I^{(2)}\propto\exp\left[-\frac{4\pi}{g^2}\left(m_W\epsilon+T
\ln\left(\frac{{\rm e}^{\gamma}}{2}c\right)-T\frac{\sqrt{2\pi}}{{\rm e}}\right)\right].$$
Therefore, the cumulant expansion converges at the temperatures obeying the inequality

$$T<\frac{m_W\epsilon}{\frac{\sqrt{2\pi}}{{\rm e}}-\ln\left(\frac{{\rm e}^{\gamma}}{2}c\right)},$$
which updates the inequality~(\ref{3}). On the other hand, since we are working in the plasma phase, i.e., 
$T\le g^2/2\pi$, it is enough to impose the following upper 
bound on the parameter of the weak-coupling approximation, $g^2/m_W$:

$$\frac{g^2}{m_W}<\frac{2\pi\epsilon}{\frac{\sqrt{2\pi}}{{\rm e}}-
\ln\left(\frac{{\rm e}^{\gamma}}{2}c\right)}.$$
Note that although this inequality is satisfied automatically at $\frac{{\rm e}^{\gamma}}{2}c\sim 1$, since  
it then takes the form $\frac{g^2}{m_W}<\sqrt{2\pi}{\rm e}\epsilon$, this is not so for the 
Bogomolny-Prasad-Sommerfield (BPS) limit,
$c\ll 1$. Indeed, in such a case,
we have $\frac{g^2}{m_W}\ln\left(\frac{2}{c{\rm e}^{\gamma}}\right)<2\pi\epsilon$, 
that owing to the logarithm is however quite feasible.

\section{Critical temperatures of the deconfining phase transition}
The deconfining phase transition in the model under study 
occurs when the density of monopoles becomes equal to the one of W-bosons. Up to inessential
subleading corrections, it takes place when the exponent of the monopole 
fugacity is equal to that of the fugacity of W-bosons~\citebk{W}. [Another way to understand why the phase transition
occurs when the two fugacities are equal to each other is to notice that once this happens,
the thickness of the string confining two W's (which is proportional to
$\xi^{-1/2}$) 
becomes equal to the average distance between the W's (proportional to
(fugacity of W's$)^{-1/2}$). 
This qualitative result was also confirmed by the RG analysis performed in Ref.~\citebk{W}.] 
The density of W's approximately equals (see Ref.~\citebk{plb} for details)
$\frac{3m_WT}{\pi}{\rm e}^{-m_W\beta}$, where we have taken into account 
that the temperatures 
of our interest are much smaller than $m_W$ in the weak-coupling regime, 
since they should not exceed $g^2/2\pi$.
Then, in the compact-QED limit, $\xi\to\beta\zeta$ 
and $T_c=\frac{g^2}{4\pi\epsilon(\infty)}$~\citebk{W}.
In the general case under discussion, $c<1$, we obtain  
the two following distinct values of critical temperatures:

\begin{equation}
\label{Tc}
T_{1,2}=g^2\epsilon\frac{1\pm\sqrt{1-\frac{b}{\pi\epsilon^2}}}{2b}.
\end{equation}
Here, $b\equiv-\frac{g^2}{m_W}\ln\left(\frac{{\rm e}^{\gamma}}{2}c\right)$, $b>0$,  
and the indices 1,2 refer to the smaller and the larger temperatures,
respectively. 
The degenerate situation $T_1=T_2=g^2/2\pi\epsilon$ then corresponds to 
$b=\pi\epsilon^2$,
and, since $\epsilon\ge 1$, $T_{1,2}\le g^2/2\pi$ in this case, as it should be. 
In particular, in the
BPS limit, $\epsilon=1$, and the deconfining phase transition takes place together
with the monopole BKT one. Obviously, at any other $b<\pi\epsilon^2$, $T_1\ne T_2$, i.e., there exist
two separate phase transitions. (Note that the existence of the 
upper bound for $b$ is quite 
natural, since in the weak-coupling regime and aside from the 
BPS limit, $b$ is definitely
bounded from above.) The existence of two phase transitions means that at $T=T_1$, 
molecules of W-bosons start dissociating, while at $T=T_2$, this process is completed. 
In another words, accounting for the interaction of monopoles via the Higgs field opens
a possibility for the existence of a new (metastable) phase at $T\in(T_1,T_2)$. This is the 
phase, where both the gas of W-molecules and W-plasma are present.

An elementary analysis shows that for 
$\pi(2\epsilon-1)<b\le\pi\epsilon^2$, $T_2<g^2/2\pi$ [and $T_2=g^2/2\pi$
at $b=\pi(2\epsilon-1)$]. At the values of $b$ lying in this interval, 
the phase transition corresponding to the critical temperature $T_2$
thus may occur. In the BPS limit, $T_2$
can only be equal to $g^2/2\pi$, that corresponds to the above-discussed case when 
both critical temperatures coincide with the one of the monopole BKT phase transition. 
In the same way, 
for any $b\le\pi\epsilon^2$, $T_1\le g^2/2\pi$, and, in particular, 
$T_1=g^2/2\pi$ only in the BPS limit, when $\epsilon=1$. Therefore, 
the phase transition corresponding to the temperature $T_1$ always takes place.
Also an elementary analysis shows that for any $\epsilon>0$ 
(and, in particular, for the realistic values $\epsilon\ge 1$)
and $b<\pi\epsilon^2$, $T_1>g^2/4\pi\epsilon$ (and consequently 
$T_2>g^2/4\pi\epsilon$ as well). Since 
$\epsilon<\epsilon(\infty)$, we conclude that both
phase transitions always occur at the temperatures which are larger than that
of the phase transition in the compact-QED limit.

Obviously, the RG analysis, performed in Ref.~\citebk{W} for the compact-QED limit remains valid, but
with the replacement $\beta\zeta\to\xi$. In particular, the deconfining
phase transition corresponds again to the IR unstable fixed point,
where the exponents of the W-fugacity $\mu\propto{\rm e}^{-m_W\beta}$ and of the monopole fugacity
$\xi$ become equal to each other [that yields the above-obtained
critical temperatures~(\ref{Tc})]. 
One can further see that the 
initial condition $\mu_{\rm in}<\xi_{\rm in}$ takes place,
provided that the initial temperature, $T_{\rm in}$, is either smaller than $T_1$ or
lies between $T_2$ and $g^2/2\pi$. For these ranges of $T_{\rm in}$,
the temperature starts decreasing according to the RG equation $dt/d\lambda=
\pi^2\bar a^4\left(\mu^2-t^2\xi^2\right)$. In this equation, $t=4\pi T/g^2$, $\lambda$ 
is the evolution parameter, $\bar a$ is some
parameter of the dimensionality [length], and 
for the comparison of $\mu$ and $\xi$ the preexponent $t^2$ is again immaterial. Then, 
in the case $T_{\rm in}<T_1$, the situation is identical to the one
discussed in Ref.~\citebk{W}, namely $\mu$ becomes irrelevant 
and decreases to zero. Indeed, from the evolution equation for $\mu$,
the following equation for $d\mu/dt$, by virtue of which one can determine the sign
of this quantity, stems:

$$\frac{d\mu}{dt}=\frac{\mu\left(2-\frac1t\right)}{\pi^2\bar a^4\left(\mu^2-t^2\xi^2\right)}.$$
One can see from this equation that if the evolution starts at $T_{\rm in}\in(g^2/8\pi,T_1)$,
$\mu$ temporaly increases until the temperature is not equal to $g^2/8\pi$, but then it nevertheless 
starts vanishing together with the temperature.
However, by virtue of the same evolution equations we see that at $T_{\rm in}\in(T_2, g^2/2\pi)$, 
the situation is now different. 
Indeed, in that case, $\mu$ is not decreasing, but rather increasing with the decrease of the 
temperature (since $d\mu/dt<0$ at $T>T_2$), until it reaches some 
value $\mu_{*}\sim{\rm e}^{-m_W/T_2}$. Once we are in the region $T<T_2$, 
the temperature starts increasing again, that together with the change of the sign of $d\mu/dt$ leads to 
the increase of $\mu$, and so on.
Thus, we see that $\mu_{*}$ is the stable local maximum of $\mu$ for such initial conditions.

\section{Including massless quarks}
Let us consider the extension of the model~(\ref{GG}) by
the fundamental dynamical quarks, which are supposed to be massless:
$\Delta S=-i\int d^3x\bar\psi\vec\gamma\vec D\psi$.
In this formula, $D_\mu\psi=\left(\partial_\mu-ig\frac{\tau^a}{2}A_\mu^a
\right)\psi$, $\bar\psi=\psi^{\dag}\beta$, where 
the Euclidean Dirac matrices are defined as $\vec\gamma=
-i\beta\vec\alpha$ with
$\beta=\left(
\begin{array}{cc}
1& 0\\
0& -1
\end{array}
\right)$ and 
$\vec\alpha=\left(
\begin{array}{cc}
0& \vec\tau\\
\vec\tau& 0
\end{array}
\right)$. Our discussion in the present Section will further follow Ref.~\citebk{plb1}.
In that paper, it has been shown that at the temperatures higher than the BKT one,
quark zero modes in the monopole field lead to the additional attraction
between a monopole and an antimonopole in the molecule. 
In particular, when the number of these modes (equal to the
number of massless flavors) is sufficiently
large, the molecule shrinks so that its size becomes of the order
of the inverse W-boson mass. Another factor which determines the size of the
molecule is the characteristic range of localization of zero modes. Namely, it can be shown that
the stronger zero modes are localized in the vicinity of the monopole center, the
smaller molecular size is. Let us consider the case when 
the Yukawa coupling of quarks with the Higgs field vanishes,
and originally massless quarks do not acquire any mass. This means that
zero modes are maximally delocalized. We shall see that in the case of one flavor, such
a weakness of the quark-mediated interaction of monopoles
opens a possibility for molecules to
undergo the phase transition
into the plasma phase at the temperature comparable with the BKT one.

It is a well known fact that in 3D, 't Hooft-Polyakov monopole is actually
an instanton~\citebk{1}. Owing to this fact, we can use the results of Ref.~\citebk{6} on the quark
contribution to the effective action of the instanton-antiinstanton molecule
in QCD. Referring the reader for details to Ref.~\citebk{plb1}, we shall present here the final expression 
for the effective action, which reads $\Gamma=2N_f\ln|a|$. Here, $a=\left<\psi_0^{\bar M}\left|g\vec\gamma\frac{\tau^a}{2}
\vec A^{a{\,}M}\right|\psi_0^M\right>$ is the matrix element of the monopole field $\vec A^{a{\,}M}$ taken between 
the zero modes $\Bigl|\psi_0^M\Bigr>$, $\Bigl|\psi_0^{\bar M}\Bigr>$
of the operator $-i\vec\gamma\vec D$ defined at the field of
a monopole and an antimonopole, respectively. The dependence of $|a|$ on the molecular size $R$ 
can be straightforwardly found and reads
$|a|\propto\int d^3r/\left(r^2\left|\vec r-\vec R\right|\right)=
-4\pi\ln(\mu R)$, where $\mu$ stands for the IR cutoff. 

At finite temperature, in the 
dimensionally-reduced
theory, the usual Coulomb interaction of monopoles~\footnote{Without
the loss of generality, we consider the molecule with the temporal component
of $\vec R$ equal to zero.}
$R^{-1}=
\sum\limits_{n=-\infty}^{+\infty}\left({\mathcal R}^2+(\beta n)^2\right)^{-1/2}$, i.e., it 
goes over into $-2T\ln(\mu{\mathcal R})$, where ${\mathcal R}$ denotes the
absolute value of the 2D vector $\vec{\mathcal R}$. 
As far as the novel logarithmic interaction,
$\ln(\mu R)= \sum\limits_{n=-\infty}^{+\infty}\ln\left[\mu
\left({\mathcal R}^2+(\beta n)^2\right)^{1/2}\right]$, is concerned, it
transforms into
$\pi T{\mathcal R}+\ln\left[1-\exp(-2\pi T{\mathcal R})\right]-\ln 2$.
Accounting for both 
interactions, we eventually arrive at the following expression for the
mean squared separation $\left<L^2\right>$ in the molecule as a function of 
$T$, $g$, and $N_f$:

$$
\left<L^2\right>=\frac{
\int\limits_{m_W^{-1}}^{\infty}d{\mathcal R}{\mathcal R}^{3-\frac{8\pi T}{g^2}}
\left[\pi T{\mathcal R}+\ln\left[1-\exp(-2\pi T{\mathcal R})\right]-\ln 2\right]^{-2N_f}}
{\int\limits_{m_W^{-1}}^{\infty}d{\mathcal R}{\mathcal R}^{1-\frac{8\pi T}{g^2}}
\left[\pi T{\mathcal R}+\ln\left[1-\exp(-2\pi T{\mathcal R})\right]-\ln 2\right]^{-2N_f}}.
$$
At large ${\mathcal R}$, $\ln 2 \ll \pi T{\mathcal R}$ and
$\bigl|\ln\left[1-\exp(-2\pi T{\mathcal R})\right]\bigr|\simeq
\exp(-2\pi T{\mathcal R})\ll
\pi T{\mathcal R}$. Hence, we see that $\left<L^2\right>$ is finite at
$T>(2-N_f)g^2/ 4\pi$, that reproduces the standard result $g^2/2\pi$
at $N_f=0$. For $N_f=1$, the plasma phase
is still present at $T<g^2/ 4\pi$,
whereas for $N_f\ge 2$ the monopole ensemble may exist only
in the molecular phase at any temperature larger than $\zeta^{1/3}$.
Clearly, at $N_f\gg \max\left\{1,{4\pi T}/{g^2}\right\}$,
$\sqrt{\left<L^2\right>}\to m_W^{-1}$,
which means that such a large number of zero modes shrinks the molecule
to the minimal admissible size.

Let us finally comment on what happens to the real deconfining phase transitions at $N_f=1$, if the 
Higgs boson is not infinitely heavy. Comparing the above-obtained critical temperatures~(\ref{Tc}) 
with the BKT critical temperature $g^2/4\pi$, we see that $T_2$ is always larger than $g^2/4\pi$, so that the second
phase transition always takes place at $T=g^2/4\pi$. As far as the first phase transition is concerned, 
one can see that $T_1<g^2/4\pi$ at $b<4\pi(\epsilon-1)$ for any $\epsilon\in[1,\epsilon(\infty)]$, and 
$T_1=g^2/4\pi$ at $b=4\pi(\epsilon-1)$.

\section{Properties of the $SU(N)$-case}
In this Section, we shall discuss some results of Ref.~\citebk{mpla} which concern the peculiarities of the $SU(N)$-case.
The $SU(N)$-generalization of the action~(\ref{1}), stemming from the $SU(N)$
Georgi-Glashow model, has the form

\begin{equation}
\label{s}
S=\int d^3x\left[\frac12(\nabla\vec\chi)^2+\frac12(\nabla\psi)^2+
\frac{m_H^2}{2}\psi^2-2\zeta{\rm e}^{g_m\psi}\sum\limits_{i}^{}
\cos\left(g_m\vec q_i\vec\chi\right)\right].
\end{equation}
Here, $\sum\limits_{i}^{}\equiv\sum\limits_{i=1}^{N(N-1)/2}$, and 
$\vec q_i$'s are the positive root vectors of the group $SU(N)$.
As well as the field $\vec\chi$, these vectors are $(N-1)$-dimensional.
Note that the $SU(3)$-version of the action~(\ref{s}), which incorporates
the effects of the Higgs field, has been discussed
in Ref.~\citebk{nd}. The compact-QED limit of the $SU(N)$-case
has been studied in Refs.~\citebk{wd} and~\citebk{sn}. The string representation
of the compact-QED limit has been explored for the $SU(3)$-case in Ref.~\citebk{epl}.
Here, similarly to all the above-mentioned papers, we assume that W-bosons corresponding to
different root vectors have the same masses.

Straightforward integration over $\psi$ then yields the following equation:

$$
S\simeq\int d^3x\left[\frac12(\nabla\vec\chi)^2-2\bar\zeta\sum\limits_{i}^{}
\cos\left(g_m\vec q_i\vec\chi\right)\right]-
$$

\begin{equation}
\label{N}
-2\bar\zeta^2\int d^3xd^3y\sum\limits_{i,j}^{}
\cos\left(g_m\vec q_i\vec\chi({\bf x})\right){\mathcal K}^{(3)}({\bf x}-{\bf y})
\cos\left(g_m\vec q_j\vec\chi({\bf y})\right).
\end{equation}
In this equation, ${\mathcal K}^{(3)}({\bf x})\equiv {\rm e}^{g_m^2D_{m_H}^{(3)}({\bf
x})}-1$, where $D_{m_H}^{(3)}({\bf x})\equiv{\rm e}^{-m_H|{\bf x}|}/(4\pi|{\bf
x}|)$ stands for the Higgs-field propagator, and
$\bar\zeta\equiv\zeta{\rm e}^{\frac{g_m^2}{2}D_{m_H}^{(3)}(0)}$
denotes the modified fugacity.
The square of the Debye mass of the field $\vec\chi$ can be derived from Eq.~(\ref{N}) by virtue
of the formula $\sum\limits_{i}^{}q_i^\alpha q_i^\beta=
\frac{N}{2}\delta^{\alpha\beta}$ and reads 
$m_D^2=g_m^2\bar\zeta N\left[1+\bar\zeta I^{(3)}N(N-1)\right]$, where $I^{(3)}\equiv\int d^3x
{\mathcal K}^{(3)}({\bf x})$. 
Note that this formula is valid also for the standard case $N=2$ and reproduces 
the $SU(3)$-result of the compact-QED limit~\citebk{epl}, \citebk{nd}
$m_D^2=3g_m^2\zeta$.

The new parameter of the cumulant expansion, $\bar\zeta I^{(3)}N(N-1)$, will be exponentially small
provided that at $x\sim 1/2$,

$$\epsilon(x)>\frac12\left[3{\rm e}^{-\sqrt{2x}}+\frac{g^2}{2\pi m_W}\ln(N(N-1))\right].$$
Setting in this inequality $x=1/2$ and recalling that~\footnote{Similarly to Ref.~\citebkcap{sn},
we assume here that the function $\epsilon$ is one and the same for any $N$.}
$\epsilon(1/2)<\epsilon(\infty)\simeq 1.787$,
we obtain the following upper bound on $N$, which is necessary (although not sufficient) to provide 
the convergence of the cumulant
expansion: $N(N-1)<{\rm e}^{15.522m_W/g^2}$. Clearly, in the weak-coupling regime under study,
this bound is exponentially large, that allows $N$ to be large enough too.

In the finite-temperature theory, owing to the fact that the root vectors have the unit length, 
the critical temperature of the monopole BKT phase transition 
remains the same as in the $SU(2)$-case, $g^2/2\pi$. In the case $m_H\sim m_W$, one can perform the RG analysis
by integrating out high-frequency modes of the fields $\vec\chi$ and $\psi$ (see the second paper of Ref.~\citebk{mpla}
for details). In particular, one can derive the leading correction to the BKT RG flow (determining the 
dependence of $\beta\zeta$ on $K$) in powers of $1/(m_Ha)^2$. Here, $a$ is the correlation radius, which 
diverges with an essential singularity at $T\to g^2/2\pi-0$ as~\footnote{In the molecular phase, the correlation 
radius becomes infinite due to the short-rangeness of the molecular fields.} $a(\tau)\sim\exp({\rm const}/\sqrt{\tau})$, 
$\tau=(g^2/2\pi-T)/(g^2/2\pi)$, while $m_H$ evolves very weakly in the critical region (namely, according to the 
equation $\frac{|dm_H^2|}{m_H^2}\ll\frac{da}{a}$) and therefore can be treated as a constant. 
In particular, from the modified RG flow
it follows that the corrections to the BKT fixed point, $K_{\rm cr.}=8\pi$, $(\beta\zeta)_{\rm cr.}=0$, vanish.
This confirms once more the result of Section 2 that the finiteness of the Higgs-boson mass does not change the critical
temperature of the BKT phase transition. 
As far as the zeroth-order [in the parameter $1/(m_Ha)^2$] part of
the free-energy density is concerned, it scales as $a^{-2}$ and therefore 
remains continuous in the critical region. Moreover, the first-order correction 
can be shown to be continuous as well.

However, an essential difference of the $SU(N)$-case, $N>2$, from the 
$SU(2)$-one does exist. Namely, while in the $SU(2)$-case
the RG invariance is exact (modulo the negligibly small high-order terms of the cumulant expansion applied
to the average over $\psi$), in the $SU(N)$-case it is only approximate, even in the compact-QED limit of the model.
In fact, in course of integration over high-frequency modes, new terms appear in the action, and the RG invariance 
holds only modulo the approximation 
$\sum\limits_{i,j}^{}{\mathcal O}_{ij}\cos\left[\left(\vec q_i-\vec q_j\right)\vec\chi{\,}\right]
\simeq\sum\limits_{i}^{}{\mathcal O}_{ii}$. Within this approximation, the RG flow of the $SU(N)$-model 
is indeed identical to the one of the $SU(2)$-case, 
since all the $N$-dependence can then be removed upon the appropriate rescaling of fields and couplings.

\section{Conclusions}
In this talk, we have discussed various properties of the (2+1)D Georgi-Glashow model at finite temperature,
in the weak-coupling limit.
First, we have explored the consequences of accounting 
for the Higgs field to the deconfining phase transition in this 
model. To this end, this field was not supposed to be infinitely
heavy, as it takes place in the compact-QED limit of the model.  
Owing to that, the Higgs field starts propagating and, in particular,
produces the additional interaction of monopoles in the plasma.
Although this effect does not change the critical temperature of the monopole BKT phase transition, it 
modifies the monopole fugacity and also
leads to the appearance of the novel terms in the action of the dual-photon field.
The cumulant expansion applied in the course of the average over the Higgs field
is checked to be convergent, provided that the weak-coupling approximation is implied in a certain sense.
Namely, the parameter of the weak-coupling approximation should be bounded from above
by a certain function of masses of the monopole, W-boson, and the Higgs field.

It has been demonstrated that although in the compact-QED limit there exists
only one critical temperature of the phase transition,
in general there exist two distinct critical temperatures.
We have discussed the dependence of these temperatures
on the parameters of the Georgi-Glashow model. In particular, 
both critical temperatures turn out to be larger than the one of the phase 
transition in the compact-QED limit. Besides that, it has been demonstrated that
the smaller of the two critical temperatures always does not exceed the critical
temperature of the monopole BKT phase transition. As far as the larger critical 
temperature is concerned, the range of parameters
of the Georgi-Glashow model has been found, where it also does not exceed the BKT one.
The situation when there exist two phase transitions implies that at the smaller 
of the two critical temperatures, W-molecules
start dissociating, while at the larger one all of them are dissociated
completely. This means that in the region of temperatures between the 
critical ones, the gas of W-molecules coexists with the W-plasma. 

From the RG equations, it follows
that the presence of the second (larger) critical temperature
leads to the appearance of a novel stable value of the W-fugacity. 
This value is reached if one starts the evolution in the region where the 
temperature is larger than the above-mentioned critical one, and the density of W's is smaller than
the one of monopoles. The resulting stable value is nonvanishing (i.e., W's at that point are still
of some importance), that is the opposite to the standard situation, which  
takes place if the evolution starts at the temperatures smaller than the first 
critical one.

Next, we have found the change of 
critical temperature of the monopole BKT phase transition
in the presence of massless dynamical quarks which interact with the Higgs boson only
via the gauge field. It has been shown that for $N_f=1$, this
temperature becomes twice smaller than the one in the
absence of quarks, whereas for $N_f\ge 2$ it becomes exponentially
small, namely of the order $\zeta^{1/3}$. The latter effect
means that this number of quark zero modes, which strengthen the attraction
of a monopole and an antimonopole in the molecule, becomes enough
for the support of the molecular phase at any temperature exceeding that
exponentially small one. Therefore, for $N_f\ge 2$, no fundamental matter
(including dynamical quarks themselves) can be confined at
such temperatures by means of the monopole mechanism. As far as the real deconfining 
phase transition at $N_f=1$ and at the finite Higgs-boson mass is concerned, 
we have seen that the larger of the two critical temperatures always exceeds the BKT one. Therefore, the second
phase transition always takes place together with the monopole BKT phase transition. At the same time, the smaller
of the two critical temperatures does not exceed the BKT one, only provided the following inequality holds:
$\frac{g^2}{m_W}\ln\left(\frac{{\rm e}^{\gamma}}{2}\frac{m_W}{m_H}\right)\le4\pi(\epsilon-1)$.

We have further investigated the general case of $SU(N)$ (2+1)D Georgi-Glashow model in the 
weak-coupling limit, at $N\ge2$.
There has been found the upper bound for $N$, which is necessary (although not sufficient) to provide
the convergence of the cumulant expansion applied in course of the average over the Higgs field. This bound is
a certain exponent of the ratio of the W-boson mass to the squared electric
coupling constant. It is therefore an exponentially large quantity in the weak-coupling
regime, that yields an enough broad range for the variation of $N$. The Debye mass of the dual photon at
arbitrary $N$ has also been found.

Finally, we have discussed 
the influence of the Higgs field to the RG flow in the $SU(N)$-version of the Georgi-Glashow model at finite temperature.
In this analysis, the Higgs-field mass was supposed to be 
large, namely of the order of the W-boson one, but not infinite, as it takes place in the compact-QED limit of the model. 
The leading correction to the coupling $g_m^2T$ then turns out to be 
inversly proportional to the second power of the correlation radius and 
therefore it vanishes at the BKT critical point. This point is one and the same for any $N\ge 2$, 
since the root vectors of the group $SU(N)$
(along which monopole charges are distributed) have the unit length. As far as the Higgs mass itself is concerned,
the RG equation for it shows that this quantity 
evolves so weakly in the vicinity of the critical point that it can be treated as a constant
with a high accuracy. Next, according to the respective RG equation, the free-energy density remains continuous
in the critical region, even with the account for the leading Higgs-inspired correction.
It has finally been mentioned that contrary to the $SU(2)$-case, in the 
$SU(N)$-model at $N>2$, the RG invariance 
holds only approximately, even in the compact-QED limit. Within this approximation, 
the RG flow in the $SU(N)$-model is identical to the one of the $SU(2)$-case. In particular, this fact confirms
the above statement that the BKT critical point is universal for any $N$.

\section*{Acknowledgments}
The author is grateful to Dr. N.O.~Agasian, in collaboration with whom 
the papers~\citebk{plb1} and~\citebk{nd} have been written, and to Profs. A.~Di~Giacomo and I.I.~Kogan for useful discussions.
This work has been supported by INFN and partially by the INTAS grant Open Call 2000, Project No. 110. 
And last but not least, the author is grateful to the organizers
of the Symposium and Workshop 
``Continuous Advances in QCD 2002/ARKADYFEST (honoring the 60th birthday of Prof. Arkady Vainshtein)'' 
(Theoretical Physics Institute of the University of Minnesota, Minneapolis, 17-23 May 2002)
for an opportunity to present these results in a very pleasant and stimulating 
atmosphere.

\end{document}